\documentclass[twocolumn, aps, prb, showpacs, preprintnumbers]{revtex4-2}
\usepackage[T1]{fontenc}
\usepackage{amsmath}
\usepackage{amssymb}
\usepackage{amsmath}
\usepackage{graphicx}
\usepackage{braket}
\usepackage{physics}
\usepackage{bm}
\usepackage{bbold}
\usepackage{mathtools}
\usepackage[utf8]{inputenc}
\usepackage{color}
\usepackage[colorlinks=true,citecolor=blue,urlcolor=black]{hyperref}

\usepackage{graphicx}% Include figure files
\usepackage{dcolumn}% Align table columns on decimal point
\usepackage[utf8]{inputenc}
\usepackage[T1]{fontenc}
\usepackage{color}

\begin{document}

\title{Optimizing quantum transport in multi-barrier graphene systems using differential evolution}

\author{Leon Browne}
\author{Stephen R. Power}%

\email{leon.browne26@mail.dcu.ie}
\email{stephen.r.power@dcu.ie}
\affiliation{School of Physical Sciences, Dublin City University, Glasnevin, Dublin 9, Ireland
}%

\date{\today}

\begin{abstract}
Potential and mass barriers in graphene introduce electron scattering, modulating transmission probabilities.
Complex multi-barrier setups allow electron transmission to be controlled with high precision, but have a huge design space of possible barrier geometries.
This work presents a framework to optimize electronic transport in such systems using differential evolution algorithms. 
First, transfer matrix methods are employed to efficiently compute quantum transport through multi-barrier structures, before optimization is applied to find barrier configurations whose transmission profiles closely match a predefined target profile. 
To address the trade-off between the accuracy and complexity of resulting barrier configurations, regularization techniques are incorporated into the optimization process. 
Our approach demonstrates the potential for highly tunable electronic transport in graphene-based systems by exploiting evolution-inspired optimization techniques.
\end{abstract}

\maketitle

\section{Introduction}
Due to its remarkable mechanical and electronic properties, graphene has been extensively investigated as a promising material for use in cutting-edge electronic devices \cite{geim2007rise,castro2009electronic, xu2024graphene}. 
Its versatility has led to applications that range from bioengineering, to flexible display technology, and quantum computing \cite{joe2022graphene, you2020laser, das2018graphene, handayani2023development, ramirez2021role}. 
However, the lack of a band gap in monolayer graphene impedes confinement and backscattering \cite{molitor2011electronic, ando1998berry}, a requirement for many devices predicated on controlling electron flow \cite{weiss2012graphene}.
An alternative approach is to exploit electrostatic gates to scatter or deflect electronic currents instead of quenching them entirely. 
However the Klein tunneling effect in graphene~\cite{katsnelson2006chiral, young2009quantum}, whereby electrons at normal incidence are transmitted with 100\% probability, places limits on the tunability of the electronic transmission in graphene using potential barriers.

Another route to overcome the limitations imposed by Klein tunneling is to use a potential which breaks the symmetry between the two sublattices in graphene \cite{lima2016controlling, aktor2021valley}. 
This can be achieved, for example, if graphene is placed on a substrate such as hexagonal boron nitride (h-BN) \cite{dean2010boron, dean2012graphene, yang2013epitaxial}.
Locally the mismatch between the lattice constants shifts the onsite energies of the two sublattices by different amounts.
This manifests as a deviation from the familiar linear dispersion found in monolayer graphene to a parabolic dispersion and the opening of a band gap~\cite{jung2015origin, giovannetti2007substrate}.
The resulting potential felt by the electrons is referred to as a \emph{mass} barrier, since the Dirac electrons acquire an effective mass due to the broken sublattice symmetry. 

Simple barrier configurations in graphene have been extensively studied with well established analytical expressions for their transmission probabilities available in the literature \cite{katsnelson2006chiral, pereira2010klein, allain2011klein}. 
However, these configurations offer a limited degree of control over transmission features. 
In contrast, complex multi-barrier configurations allow for much greater tunability by nature. 
These can be applied using patterned electrodes~\cite{young2009quantum, marconcini2019geometry, rickhaus2015guiding, wang2019graphene} or substrate design \cite{hinnefeld2018graphene, forsythe2018band, barcons2022engineering,  kammarchedu2024understanding, lassaline2025gradient} to allow the spatial distribution of potentials to be controlled.
In practice, imposing complex mass barrier configurations in graphene is more difficult to achieve than applying potentials.
For graphene on h-BN, the strength of the mass could be controlled by varying the interlayer separation using pressure, \cite{yankowitz2016pressure, yankowitz2019tuning, fulop2021boosting} or by applying strain to manipulate the local stacking registry \cite{yang2023heterostrain, khatibi2019impacts}.
However, the spatially varying band gaps induced by mass barriers have a more feasible analogue in Bernal-stacked bilayer graphene (BLG), where, dual gating allows gaps to be opened using a layer, rather than a sublattice, asymmetry ~\cite{mccann2013electronic,zhan2012engineering,zhang2009direct}. This once more allows patterned electrodes or substrates to modulate the mass felt by electrons in different regions of a device.
%Here, dual gating introduces different potentials on each layer, instead of sublattice.

While arbitrary barrier geometries allow for a more precise tuning of electronic transport, the absence of a simple analytic expression makes it more difficult both to calculate the expected transmission for a specific barrier configuration and, conversely, to find the barrier configuration required for a desired electronic behavior.
To overcome the former issue, the transfer matrix (TM) method~\cite{mackay2022transfer} can be used to efficiently calculate the transmission through a complex multi-barrier system numerically~\cite{xu2015transmission, barbier2009bilayer, seffadi2025effect, wang2010electronic, zalipaev2015resonant, pellegrino2011transport,
nguyen2016transfer, dell2009multiple, ghosh2009electron, grover2012transfer, lima2018tuning,
phan2021electronic, li2009generalized, wang2014transfer, wang2010electronic, rodriguez2012resonant}. In this work, we demonstrate that differential evolution~\cite{storn1997differential, bujok2021real, yang2021crystal, ma2015differential, yang2019improved, zahedinejad2014evolutionary}, an optimization procedure which mimics natural evolution, can be used to solve the second problem and find barrier geometries whose transmission properties converge to closely match arbitrary target profiles. 
For the one-dimensional barrier systems considered in this work, this allows the geometries required for arbitrary band-pass and band-stop filters to be found, as well as geometries allowing highly customized energy-dependent transmissions.  
Combined with recent experimental progress, where arbitrary potential profiles can be created in graphene by precise lithographic control of the thickness of the encapsulating h-BN~\cite{lassaline2025gradient}, our method enables the reverse engineering of functional electronic devices from their desired transport characteristics.

We investigate how the results obtained with our approach depend on a number of variables, including the choice of evolutionary parameters and the number of individual electrostatic regions in the scattering region.
Given the experimental difficulty in realizing extremely complicated barrier structures, we examine how the trade-off between the complexity of the barrier configuration and the precision of the resulting electronic transmission can be controlled by adding a regularization term to the optimization process \cite{gerth2019regularized}.
This can be used to penalize systems which give a very precise match with the target electronic profiles, but which would be very difficult to achieve in experiment. 
We finish by demonstrating that the approach can be generalized in two further ways: first to optimize the angular, instead of the energy, dependence of the transmission, and secondly to optimize angle-averaged transmission, reflecting the spread of incident angles to be expected in a typical two-terminal electronic measurement.

\begin{figure} 
    \centering
    \includegraphics[width=\linewidth]{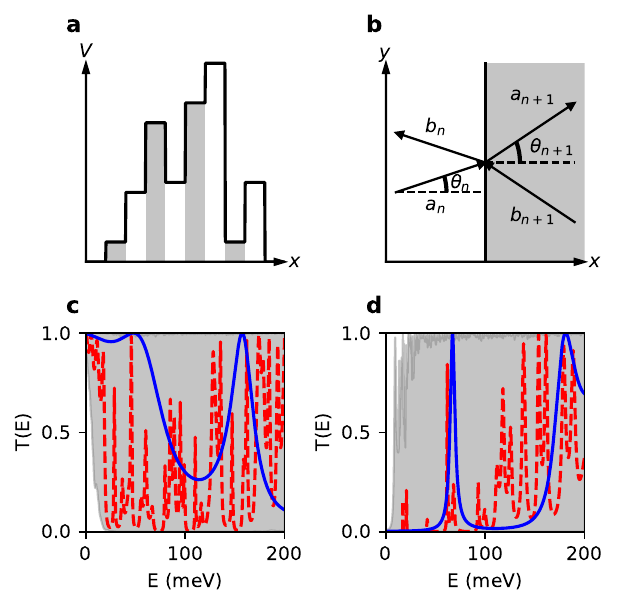}
    \caption{(a) Schematic of an arbitrary barrier configuration which can be considered using the transfer matrix method (b) Coefficients of right- $(a_n)$ and left-going $(b_n)$ waves on either side of an interface. Single angle transmissions through multi-barrier (c) potential and (d) mass systems where $\theta_\mathrm{inc.}=0.3$ rads and $\theta_\mathrm{inc.}=0$ rads, respectively. The blue curves in these panels correspond to simple double barrier systems, whereas the red curves show results for systems with 50 random barriers. The shaded regions highlight the domains spanned by 2000 such random configurations. The maximum barrier strength is set at $V_P=200$ meV for potential and $\Delta_M=240$ meV for mass barriers.} 
       \label{fig1}
\end{figure}

\section{Transfer Matrix Method}
We begin by outlining the TM method to calculate the transmission through an arbitrary one-dimensional barrier system in graphene, such as that shown in Fig.~\ref{fig1}(a).
The Hamiltonian in the $n^\mathrm{th}$ region can be written 
\begin{equation}
\mathcal{H}^{(n)}=
\hbar v_f \begin{pmatrix}
V_A^{(n)}&k_x^{(n)}-ik_y\\
k_x^{(n)}+ik_y&V_B^{(n)}
\end{pmatrix},
\label{eq:1}
\end{equation}
where $V_A^{(n)}$ and $V_B^{(n)}$ denote the potentials on each of the two sublattices.
In the case of an electrostatic potential barrier we set $V_A^{(n)} = V_B^{(n)} \equiv V^{(n)}$, whereas in a mass barrier $V_A^{(n)} = -V_B^{(n)} \equiv \frac{\Delta^{(n)}}{2}$, where the mass term $\Delta^{(n)}$ is the magnitude of the band gap opened.
In this work, we will assume the complete system to have only potential or mass barriers, but our approach can easily be extended to more general cases where both components are present.
The $y$ component of the wavevector is conserved, so that $k_y^{(n)} \equiv k_y$ for all regions $n$.
The wavevector magnitude $k^{(n)}=\sqrt{(k_x^{(n)})^2 + (k_y^{(n)})^2}$ and hence $k_x^{(n)}$ in each region is found by solving the Schrodinger equation.
The total wavefunction in each region can be written
\begin{multline}
    \Psi^{(n)} =  \frac{a_n}{\sqrt{2}} 
    \begin{pmatrix}
    C_A^{(n)}  \\
    \eta_n e^{i \theta_n} C_B^{(n)}  
    \end{pmatrix} e^{i \left(k^{(n)}_x x + k_y y \right)} \qquad 
    \\  +
     \frac{b_n}{\sqrt{2}} 
    \begin{pmatrix}
    C_A^{(n)} \\
    \eta_n e^{i ( \pi -\theta_n)} C_B^{(n)}  
    \end{pmatrix} e^{i \left(- k^{(n)}_x x + k_y y \right)} \,,
    \label{wavefn}
\end{multline}
where the two terms weighted by coefficients $a_n$ and $b_n$ correspond to waves propagating from the left and right, respectively, as shown in Fig.~\ref{fig1}(b).
The coefficients $C_A^{(n)}$, $C_B^{(n)}$ depend on the relative values of $E$, $V_A$ and $V_B$
\begin{align}
C_A^{(n)}(E) &= \sqrt{\frac{2|E - V_B^{(n)}|}{|E - V_A^{(n)}| + |E - V_B^{(n)}|}} \\
C_B^{(n)}(E) &= \sqrt{\frac{2|E - V_A^{(n)}|}{|E - V_A^{(n)}| + |E - V_B^{(n)}|}},
\label{general_C_AB}
\end{align}
where $E$ is the electron energy. The factor $\eta_n$, which represents a generalized band index, also depends on these values \cite{aktor2021valley}
\begin{equation}
\eta_{AB} =
\begin{cases}
1 & \text{for } V_A,\, V_B < E \\[4pt]
-1 & \text{for } V_A,\, V_B > E \\[4pt]
i & \text{for } V_B < E < V_A \\[4pt]
-i & \text{for } V_A < E < V_B
\end{cases}\,.
\label{general_band_index}
\end{equation}
In the case of an electrostatic potential, these quantities reduce to $C_A^{(n)}=C_B^{(n)}=1$ and $\eta_n = \mathrm{sign} (E - V^{(n)})$.
The transmission through a barrier, or system of barriers, can be found by enforcing wavefunction continuity, $\Psi^{(n)} = \Psi^{(n+1)}$ at each interface in the system. 
Each interface gives rise to two equations in terms of the unknown wavefunction coefficients $a_n, b_n$, and the resultant set of simultaneous equations can be solved to obtain these. 
For a small number of interfaces, this allows an analytic expression to be derived for the transmission probability $T = 1 - b_0 b_0^*$.
However, the number of equations increases with the number of interfaces, making the analytic approach cumbersome for the arbitrary barrier configurations of interest here.
The TM method offers a more efficient approach, by relating the wavefunction amplitudes at either side of the interface in Fig.~\ref{fig1}(b) using the equation
\begin{equation}
\begin{pmatrix} a_{n+1} \\ b_{n+1} \end{pmatrix} = 
\mathbf{M}_n
\begin{pmatrix} a_n \\ b_n \end{pmatrix},
\label{eq:singleM}
\end{equation}
where $\mathbf{M}_n$ is the $2\times2$ transfer matrix of the $n^\mathrm{th}$ interface.
This can be found for a single interface by rearranging the simultaneous equations resulting from wavefunction continuity, and using Eq.~\eqref{wavefn} we find 
\begin{widetext}
\begin{equation}
\mathbf{M}_n = \frac{1}{2\eta_{n+1}C_A^{(n+1)}C_B^{(n+1)}\cos(\theta_{n+1})} 
\left(
\begin{array}{c c}
\substack{
\left(\eta_{n+1}C_A^{(n)}C_B^{(n+1)} e^{-i\theta_{n+1}} + \eta_n C_A^{(n+1)}C_B^{(n)} e^{i\theta_n}\right) \\
\times e^{i\left(-k^{(n+1)}_x + k^{(n)}_x\right)x}
}
& 
\substack{
\left(\eta_{n+1}C_A^{(n)}C_B^{(n+1)} e^{-i\theta_{n+1}} - \eta_n C_A^{(n+1)}C_B^{(n)} e^{-i\theta_n}\right) \\
\times e^{i\left(-k^{(n+1)}_x - k^{(n)}_x\right)x}
}
\\[1em]
\substack{
\left(\eta_{n+1}C_A^{(n)}C_B^{(n+1)} e^{i\theta_{n+1}} - \eta_n C_A^{(n+1)}C_B^{(n)} e^{i\theta_n}\right) \\
\times e^{i\left(k^{(n+1)}_x + k^{(n)}_x\right)x}
}
&
\substack{
\left(\eta_{n+1}C_A^{(n)}C_B^{(n+1)} e^{i\theta_{n+1}} + \eta_n C_A^{(n+1)}C_B^{(n)} e^{-i\theta_n}\right) \\
\times e^{i\left(k^{(n+1)}_x - k^{(n)}_x\right)x}
}
\end{array}
\right).
\label{eq:4}
\end{equation}
\end{widetext}
For a system with $N$ barrier segments ($N+1$ interfaces), Eq.~\eqref{eq:singleM} can be applied repeatedly to obtain
\begin{equation}
\begin{pmatrix} a_N \\ b_N \end{pmatrix} = 
\mathbf{M} 
\begin{pmatrix} a_1 \\ b_1 \end{pmatrix},
\label{eq:5}
\end{equation}
where $\mathbf{M} = \mathbf{M}_{N+1}\mathbf{M}_{N} \dots \mathbf{M}_2 \mathbf{M}_1$ is the transfer matrix of the entire system.
Once $\mathbf{M}$ is known, the transmission probability of an electron through the barrier configuration is calculated from its elements as follows~\cite{markos2008wave}
\begin{equation}
T=1-\left(\frac{M_{12}}{M_{22}}\right)^*\left(\frac{M_{12}}{M_{22}}\right) \,.
\label{eq:6}
\end{equation}
The TM approach therefore reduces the calculation of transmission to repeated $2\times2$ matrix multiplication, allowing for arbitrarily complex barrier systems to be efficiently evaluated. 
The TM approach can alternatively be written within a finite difference representation \cite{tworzydlo2008finite}, which can give rise to fermion doubling effects under certain circumstances. In our case, local analytic solutions are instead used in each barrier region, which avoids this problem.
The TM method has been used to investigate a wide range of properties in graphene, using systems of multiple patterned barriers~\cite{xu2015transmission, barbier2009bilayer, seffadi2025effect, wang2010electronic}, realistic smooth barriers~\cite{zalipaev2015resonant, pellegrino2011transport} and circular quantum dot barriers~\cite{nguyen2016transfer}. Some of these properties involve magnetic field barriers~\cite{dell2009multiple, ghosh2009electron, grover2012transfer, park2019:magnetoelectrically}, Fermi velocity engineering~\cite{lima2018tuning, phan2021electronic} and strain effects~\cite{pellegrino2011transport, phan2021electronic}. 

Figs.~\ref{fig1}(c) and (d) illustrate the energy-dependent transmissions at fixed incident angles $\theta_\mathrm{inc.}=0.3$ rads and $\theta_\mathrm{inc.}=0$ for potential and mass barrier systems, respectively.
We consider a non-normal incident angle for the potential barrier case, as Klein tunneling otherwise gives $T=1$ for all energies and barrier heights. The blue lines in each case show the transmission through a simple configuration of two identical ($V_P=200$ meV for potential, $\Delta_M=240$ meV for mass) barriers, where the barrier width and separation are both $10$ nm.
These systems exhibit familiar and predictable behavior, including clear Fabry-P\'erot resonances in the potential barrier system arising from the constructive and destructive interference between the forward and back-scattered electron waves at each interface. 
The mass barrier has a low transmission, apart from a single resonance peak, until the electron energy is above the band gap introduced in the barrier regions by the mass term. 
In contrast to these simple behaviors, the dashed red lines in each panel show the transmission for a disordered system with 50 barrier regions, each of 5 nm width, where each barrier magnitude is chosen randomly between $0$ and $V_P$ or $\Delta_M$ as appropriate.
The role of barrier width, and its connection with experimental techniques, is discussed in more detail in Sec. \ref{sec:discussion}.
The gray shaded areas represent the ranges spanned by 2,000 such randomly generated configurations for each system.
The fine level of detail seen for individual configurations, together with the wide range of possible solutions, suggests that designer transmissions can be precisely realized once suitable barrier configurations are known.

\section{Differential Evolution}
To find barrier configurations for specific transport behaviors, we use differential evolution (DE), an optimization strategy inspired by biological evolution, which iteratively refines a population of possible solutions over multiple generations in order to maximize a fitness function. 
DE algorithms excel at minimization and have been successfully applied to engineering problems \cite{bujok2021real} and physics problems such as crystal structure prediction \cite{yang2021crystal}, quantum control \cite{ma2015differential, yang2019improved, zahedinejad2014evolutionary} and electron optics~\cite{ildarabadi2025tunable}.
The general steps in an evolutionary algorithm are: (i) the \emph{encoding} of a possible solution into a `chromosome' made of constituent `genes'; (ii) the \emph{initialization} of a randomly generated population; (iii) the \emph{evaluation} of the fitness of each individual in the population; (iv) the \emph{selection} of parent individuals from the population, with a preference given to those with higher fitness; this phase includes gene mutation and crossover; (v) the \emph{reproduction} of parent individuals to form offspring and populate the next generation of solutions.
Phases (iii)-(v) are repeated until the fitness converges.
We note that, in practice, we choose to minimize a \emph{loss} function measuring the unsuitability of a solution instead of maximizing the fitness.

In this work, we encode configurations of $N$ equal width barriers using $N$-dimensional vectors (chromosomes), with each dimension (gene) storing the height of the corresponding barrier. 
The initial parent population is randomly generated, and for a given target transmission profile, the loss of each individual is calculated using the mean absolute error (MAE) over a chosen energy range to quantify the difference between the target and individual transmissions.
For each parent vector, $\mathbf{x}_i$, in the population, a corresponding \emph{mutated} vector $\mathbf{u}_i$ is created using the equation~\cite{engelbrecht2007computational}
\begin{equation}
\mathbf{u}_i = {\mathbf{x}_i}_a + \beta\left({\mathbf{x}_i}_b-{\mathbf{x}_i}_c\right),
\label{mutated_v}
\end{equation}
where ${\mathbf{x}_i}_a$, ${\mathbf{x}_i}_b$ and ${\mathbf{x}_i}_c$ are three randomly selected individuals from the current population, not including  $\mathbf{x}_i$.
$\mathbf{u}_i$ can be understood as a mutated version of a randomly chosen parent vector, where the direction of the mutation is determined by the difference vector between two more randomly chosen parent vectors, scaled by the mutation rate $\beta$.
The chromosome of a potential offspring solution, or \emph{trial} vector, is generated during the crossover phase using those of the parent and mutated vectors.
This is achieved by probabilistically replacing each gene (\emph{i.e} strength of a particular barrier segment) from the parent vector with that of the mutated vector according to the crossover rate, $C$.
The fitness of the offspring solution, which contains genes from both $\mathbf{u}_i$ and $\mathbf{x}_i$, is then compared to its parent, and the parent solution is replaced if the offspring is fitter. 

As the algorithm converges to a solution over several generations, the difference vectors become smaller in size, as individuals inside the population contain increasingly similar genes. 
The larger the value of $\beta$, the more exploratory the algorithm will be in solution space, since the mutation rate directly controls how different a mutated vector will be from an existing solution. 
The higher values of $C$, meanwhile, determine the strength of gene transfer from the mutated vector to the offspring vector, giving greater weight to the genes found in exploration. 
As we show below, there are different strategies to vary these parameters between generations and which affect the efficiency of the optimization.

\begin{figure}
    \centering
    \includegraphics[width=\linewidth]{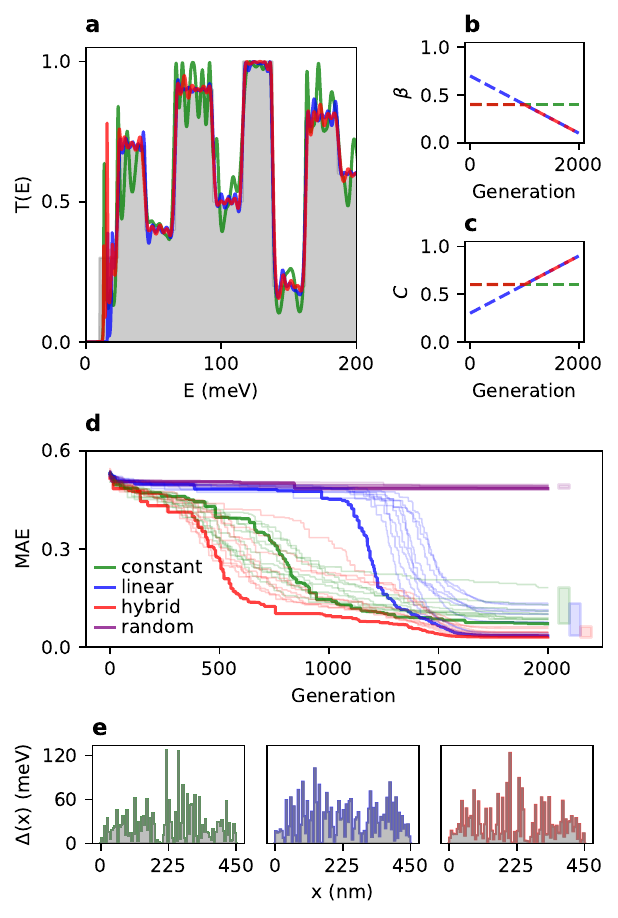}
    \caption{\textbf{Optimized mass barriers at normal incidence ($\theta_\mathrm{inc.}=0$)} (a) An arbitrary target transmission profile (gray shaded area), with colored lines showing optimized transmissions for the different mutation ($\beta$) and crossover ($C$) parameter strategies in panels (b) and (c).
    (d) The evolution of the mean absolute error (MAE) over 2000 generations for 10 runs of each strategy, with the best solution for each case shown in bold. The purple lines correspond to a purely random strategy. The shaded boxes show the range of results obtained for each strategy.
    (e) Optimized barrier configurations for each strategy.}
    \label{fig2}
\end{figure}

\section{Implementation and Optimization Strategies}
A mass barrier system with normally incident electrons is selected to demonstrate the approach, as the absence of Klein tunneling gives a greater range of possible behaviors in comparison to pure potential barriers. 
All the results~\cite{zenodo_data} that follow consider barrier segments that are $5$ nm wide, but the basic approach can be easily rescaled.
A complicated piecewise transmission, shown by the gray shaded area in Fig.~\ref{fig2}(a), was chosen to act as the target transmission.   
This panel also shows the resulting transmissions for optimized structures generated using different strategies to vary the mutation ($\beta$) and crossover ($C$) rates, as shown in Fig.~\ref{fig2}(b,c).
For all the results shown here, we consider $N=90$ barriers and a fixed population size of $P=100$ individual solutions. 
By performing multiple iterations with each set of parameters, we can account for variability due to stochastic effects. 
Fig.~\ref{fig2}(d) shows, at each generation, the mean absolute error (MAE) of
10 independent trials for each strategy, with the best-performing result for each shown in a thicker line.

The green curves in each panel of Fig.~\ref{fig2} correspond to the simplest strategy: fixed values of $\beta = 0.4, C = 0.6$ across all generations.
The corresponding `best-performing' optimization, shown by the solid green curve in Fig.~\ref{fig2}(d), shows how the MAE between the target and best individual solution decreased steadily during the DE optimization, with the final optimal barrier configuration shown in green in Fig.~\ref{fig2}(e).
The green shaded box in Fig.~\ref{fig2}(d) shows the range of MAE values obtained by different iterations of the optimization using the same parameters.
To demonstrate that the DE procedure is indeed efficiently searching the solution space, Fig.~\ref{fig2}(d) also includes evolution curves (shown in purple) for a strategy with no optimization.
This strategy simply generates a population of completely random individuals at each generation.
The DE approach far outperforms the random strategy, which is extremely inefficient at finding a suitable solution.

The blue curves in Fig.~\ref{fig2} show the corresponding results for a strategy where $\beta$ and $C$ are linearly varied during optimization. 
The mutation rate is initially set at a high value ($\beta=0.6$) and then decreased to $\beta=0.1$, while the crossover rate is increased to $C=0.9$ from a smaller initial value ($C=0.4$).
Larger values of $\beta$ force the algorithm to explore the solution space more thoroughly early in the optimization process by creating mutation vectors, using Eq.~\eqref{mutated_v}, which are very different from the existing population. 
As the optimization proceeds and the algorithm converges towards a solution, the difference vectors entering in Eq.~\eqref{mutated_v}  become smaller as individuals inside the population contain increasingly similar genes.
Decreasing $\beta$ and increasing $C$ at this stage shifts the emphasis from mutation to gene swapping, so that new solutions are created by combining existing genes.
The shift of emphasis from exploration to fine-tuning as the optimization proceeds is somewhat similar to the role of decreasing temperature in simulated annealing \cite{kirkpatrick1983optimization}.

Comparing the linear (blue) and constant (green) parameter strategies in Fig.~\ref{fig2}(a) and (d), we see that the linear strategy gives considerably better final results, while the constant parameter strategy seems not to have fully converged. 
However, while the constant strategy shows a steady improvement across generations, the linear model is much slower in earlier generations. 
Indeed, it is barely an improvement upon the completely random model for the first 1000 generations, and it takes almost 1500 generations to eventually surpass the constant strategy. 
To try and combine the strong early performance of the constant model with the more accurate later convergence of the linear model, we developed a hybrid strategy shown by the red dashed lines in  Fig.~\ref{fig2}(b) and (c).
This approach uses the constant $\beta$ and $C$ values for the first half of the optimization, before linearly varying them for the final stages.
As shown in Fig.~\ref{fig2}(d), this approach indeed combines the strengths of the individual strategies, giving a slightly fitter overall solution than the linear strategy but also giving an excellent performance across all stages of the evolution. 
Thus, for the remainder of this work, the hybrid tuning will be employed to control the evolutionary rates.

\begin{figure}
    \centering
\includegraphics[width=\linewidth]{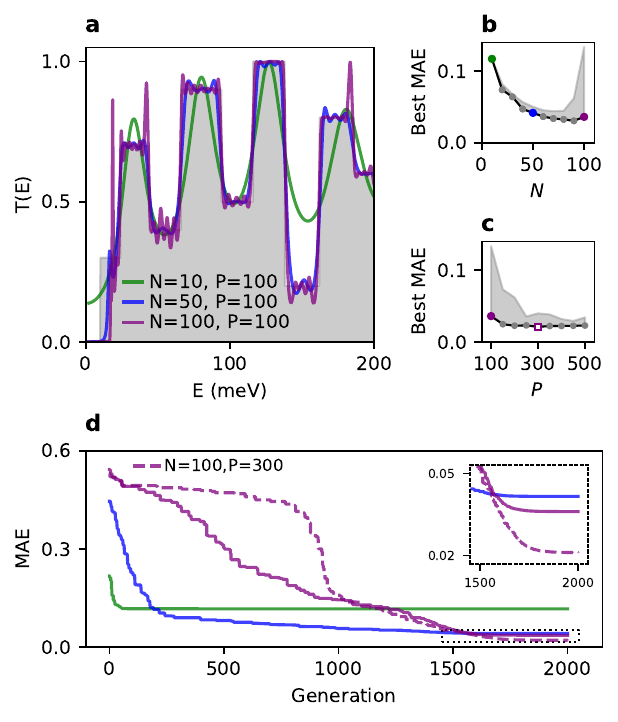}
    \caption{\textbf{Effect of changing barrier number ($N$) and population size ($P$) for mass barriers at normal incidence} 
    (a) An arbitrary target transmission profile (gray shaded area), with colored lines showing optimized transmissions for different values of $N$. 
    (b) Accuracy achieved after 2,000 generations using different numbers of barriers ($N$) with a fixed population size ($P=100$). Colored dots correspond to the cases shown in panel (a).
    (c) Accuracy achieved for different population sizes $P$, with a fixed number of barriers $N=100$. 
    (d) The evolution of the MAE over 2000 generations for different cases in (b) and (c), with the inset showing a zoom near the final convergence.}
    \label{fig3}
\end{figure}

The transmission of the converged optimal systems in Fig.~\ref{fig2} is in excellent agreement with the target transmission profile, but this comes with two significant costs: very complex barrier configurations and lengthy optimization procedures.
We now examine the three-way tradeoff between accuracy, complexity and convergence efficiency by comparing optimizations with different numbers of barriers or population sizes. 
Fig.~\ref{fig3}(a) shows transmissions for optimized systems of $N=10, 50, 100$ barriers, with Fig.~\ref{fig3}(b), showing how the best accuracy (MAE) achieved varies as the number of barriers is increased. 
The colored symbols in Fig.~\ref{fig3}(b) correspond to the transmissions shown in panel (a).
The shaded region in Fig.~\ref{fig3}(b) shows the range of outcomes achieved using 10 independent trials for each value of $N$, with a fixed population size $P=100$.
The accuracy of the solutions achieved can decrease for both small and large values of $N$.
Systems with small numbers of barriers have less room for optimization due to their limited tunability, and therefore perform poorly in comparison to larger systems. 
However, beyond $N=80$, many of the optimized structures have transmissions which deviate considerably from the target, as is evident by the wider spread of MAE values in Fig.~\ref{fig3}(b).
As $N$ increases, the algorithm struggles to find suitable solutions within the rapidly growing search space, and therefore more resources, in the form of a larger population or number of generations, may be required to reliably optimize systems of this size. 

Fig.~\ref{fig3}(c) shows the effect of increasing the population size for the $N=100$ case, which had the widest range of accuracies in Fig.~\ref{fig3}(b). 
The solid purple point in both panels represent the same result. 
As we consider larger values of $P$ in Fig.~\ref{fig3}(c), we see a reduction in both the best MAE achieved (shown by the symbols) and the range of results achieved by different iterations (shaded area), suggesting that the algorithm now has more opportunities to find better solutions and fully converge.
This is confirmed by the evolution curves plotted in Fig.~\ref{fig3}(d).
The solid curves show the convergence of the MAE for the different size systems in Fig.~\ref{fig3}(a) with a fixed population $P=100$.
Larger systems take longer to converge, but ultimately give a better match with the target transmission. 
The dashed purple line shows the convergence for a system with $N=100$ barriers and $P=300$, corresponding to the hollow purple dot in Fig.~\ref{fig3}(c).
Comparing the two purple curves, we see that a larger population size leads to a considerably slower convergence, particularly in the early stages of optimization, but eventually gives a better fit with a final MAE of $\sim 0.02$. 
While increasing the population size or the number of generations increases the computational complexity of the optimizations, larger numbers of barriers $N$ also lead to more complex structures which will be more difficult to realize in experiment.
Furthermore, beyond $N=50$ the improvements are minimal, resulting in a smaller return on investment with a rapidly increasing level of complexity.
We therefore take $N=50$ to be a good balance between accuracy and complexity, and use it as the maximum number of barrier regions for the remaining results. 
 
\section{Regularization}
It is not only the number of barriers, but also the distribution of potentials, which could result in complex potential profiles that are difficult to realize experimentally. 
Given the large number of barriers in these systems, individually contacted gates are not particularly feasible and a more realistic implementation could involve substrate engineering to control the spatial distribution of the potential induced by a single gate~\cite{hinnefeld2018graphene, forsythe2018band, barcons2022engineering, lassaline2025gradient}. 
Substrate engineering is more suited to more smoothly varying potential profiles, however, and could struggle with the optimized barrier configurations generated so far, such as those in Fig.~\ref{fig2}(e), which contain a large number of sharp discontinuities.
To account for this, we now consider the role of regularization -- a technique often employed in machine learning to discourage an algorithm from generating overly complex results \cite{gerth2019regularized}. 

In our case, we define the \emph{complexity} of a barrier configuration as 
\begin{equation}
    \text{Complexity} = \frac{1}{N-1}\sum_{i=1}^{N-1} \left| V_{i+1} - V_{i} \right| \,.
\label{eq:8}
\end{equation}
This measures the average height difference between consecutive barriers in a configuration.
To penalize complex configurations, we modify the loss function to include a regularization term proportional to the complexity 
\begin{equation}
    \text{Loss} = \text{MAE} + \alpha \times \text{Complexity} \,,
\label{Loss}
\end{equation}
where the regularization coefficient $\alpha$ controls the relative importance of achieving a good fit with the target transmission and reducing the complexity of the final barrier system. 

\begin{figure}
    \centering
    \includegraphics[width=\linewidth]{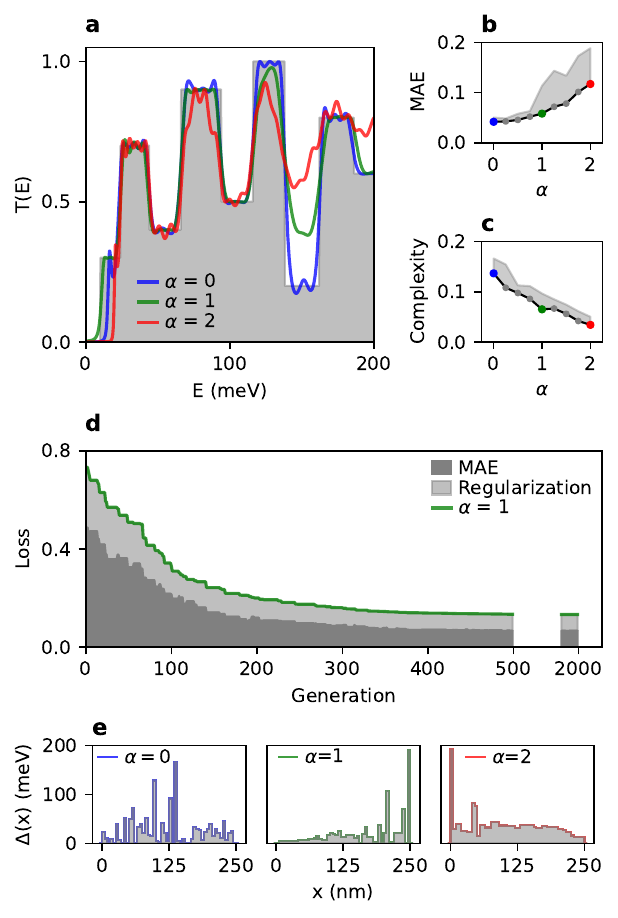}
    \caption{\textbf{Effect of regularization for mass barriers with $\theta_\mathrm{inc.}=0$} (a) An arbitrary target transmission profile (gray shaded area), with colored lines showing optimized transmissions for different regularization strengths ($\alpha$).
    (b) The MAE component of the optimized loss found for systems of various $\alpha$ values. (c) The barrier complexity associated with each optimized configuration. (d) The evolution of the MAE over 2000 generations for $\alpha=1$. (e) The best-fit barrier configurations for different $\alpha$ values.
        }
    \label{fig4}
\end{figure}

Fig.~\ref{fig4} illustrates the effect of regularization on an $N=50$ barrier system optimized over 2000 generations with $P=100$.
Panel (a) compares the target and optimized transmissions for $\alpha=0,1,2$. 
The unregularized data (blue) in this figure is the same as that presented for $N=50$ in Fig.~\ref{fig3}. 
As regularization increases, the optimized transmissions fail to capture some of the fine details of the target transmission profile.   
This reduction in accuracy is more pronounced for both higher energies and larger values of $\alpha$, which can be seen by the red line ($\alpha=2$) failing to capture the desired transmission dip at higher energies. 
However, even with such large regularization, the transmission is still an excellent match for the complex target over a wide energy range. 

The increasing mismatch between the target and optimized transmission profiles is shown quantitatively in Fig.~\ref{fig4}(b), which plots the MAE portion of the best-fit regularized solution as a function of the regularization parameter $\alpha$, with the colored symbols corresponding to the cases shown in Fig.~\ref{fig4}(a). 
The shaded region again represents the range of results obtained from 10 independent trials, this time for each value of $\alpha$.
As we increase the regularization strength, the MAE also steadily increases, as does the width of the shaded region.
These increases follow from the growing focus on achieving simpler, rather than more accurate, barrier configurations.
In the limit of very large $\alpha$, the model would return the simplest possible barrier structure with every segment having the same potential value, which would depend on the exact optimization trajectory and take a different, random value for each iteration.
The corresponding reduction in the complexity of the optimized structures for larger $\alpha$ is shown similarly in Fig.~\ref{fig4}(c). 
As expected, this quantity decreases steadily as the penalty for complex barrier structures increases.

Fig.~\ref{fig4}(d) shows that, with moderate levels of regularization, the costs associated with accuracy and complexity decrease in tandem throughout the optimization process.
The dark and light shaded regions in this plot correspond to the MAE and regularization contributions to the total loss for the best-performing $\alpha=1$ optimization. 
Finally, the optimized structures for the $\alpha=0,1,2$ cases are shown in Fig.~\ref{fig4}(e), with the structures becoming clearly more smooth as the regularization is increased.
In particular, the solution for $\alpha=2$ is slowly varying over most of the barrier region, with only a very few sharp barriers.
This could be implemented using a combination of a relatively smooth variation of the substrate height over most of the device, combined with a small number of sharper substrate features or independently contacted barriers to produce the required potential profile. 
This design approach can directly complement works such as \cite{lassaline2025gradient}, as the simpler and smoother barrier landscapes produced by our regularized model can be realistically approximated by the $20$ nm experimental pixel size. 
Regularization can clearly be used to fine-tune the balance between barrier complexity determined by experimental constraints and the transmission accuracy required for a particular application.

\section{Extensions and Discussion}
\label{sec:discussion}
So far, we have introduced the basic optimization framework and discussed various optimization strategies and regularization techniques.
We have focused on reproducing target transmissions while balancing the accuracy of the optimized transmission and the feasibility of the associated potential profile. 
The target transmission considered to date was a complicated piecewise function, whose random nature served as a stress test to demonstrate the capability of the algorithm to find and fine-tune suitable barrier configurations. 
We now consider some more realistic target behaviors and consider how the optimization approach can be further extended to other systems of interest. 

\begin{figure}
    \centering
    \includegraphics[width=\linewidth]{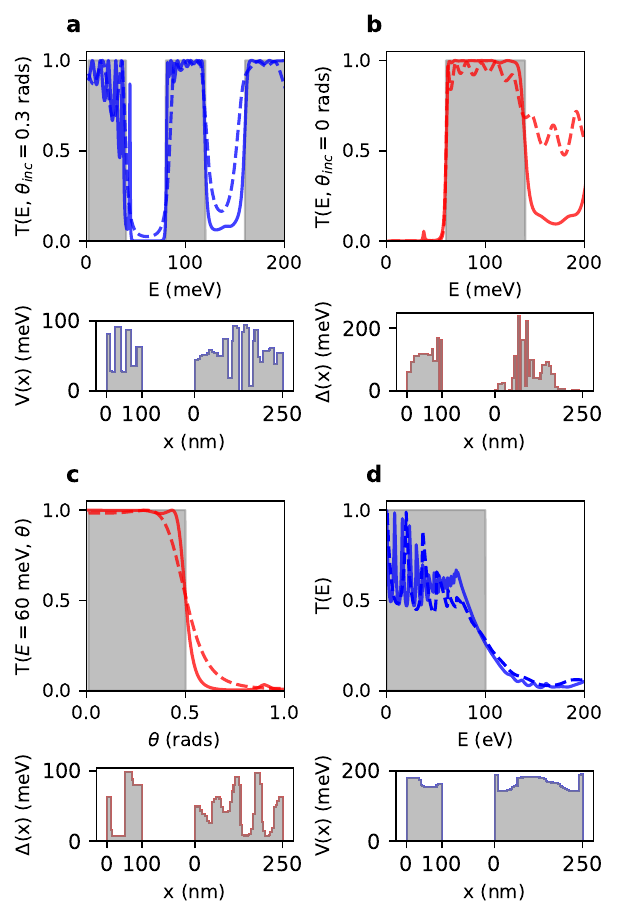}
    \caption{\textbf{Optimization for different target behaviors.} 
    In all cases, a regularization of $\alpha=1$ is applied and both short ($N=20$, dashed) and long ($N=50$, solid) barrier systems are considered. The top subpanel for each case shows the target (shaded) and optimized transmissions, and the bottom subpanel shows both optimized structures.
    (a) and (b) have targets relevant for (multi) band pass or stop filters, using potential (blue) or mass (red) barriers and incident angles $\theta_\mathrm{inc.}=0.3$ and $\theta_\mathrm{inc.}=0$ rads, respectively.
    (c) shows the optimization of an angular-dependent target, corresponding to beam collimation, for $E=60$ meV.
    (d) considers incident electrons across a range of angles using an angle-averaged transmission in the target.
    }  
    \label{fig5}
\end{figure}

Fig.~\ref{fig5} investigates the use of potential (blue) and mass barriers (red) as electron (multi) band-pass filters and collimators, with all simulations including a regularization term with $\alpha=1$.
Panels (a) and (b) consider band-pass applications, where the objective is to maximize transmission in the shaded energy regions, and minimize it elsewhere.
The dashed and solid lines show the optimized transmissions for systems with $N=20$ and $50$, respectively.
In both cases, the incident angles considered in Fig.~\ref{fig1} are employed again, and we note that the target transmissions are contained within the shaded `design spaces' from Fig.~\ref{fig1}(c) and (d).
We note that the potential barrier system does an excellent job at mimicking the desired multi band-pass behavior in Fig.~\ref{fig5}(a), which is consistent with previous studies showing that periodic barrier arrays can induce similar behavior \cite{sanchez2019non}.
Indeed, we notice that associated optimized barrier configurations, shown in the subpanel underneath, contain quite regular domains consisting of barriers with roughly similar widths, heights and separations.
The lack of perfect regularity in our structures, together with the differences between the shorter and longer optimized structures, suggests that the required electronic behavior may be reasonably robust against disorder\cite{kammarchedu2024understanding}.
However, we note that the optimization procedure could be amended by using a configurationally-averaged transmission over different disorder realizations, instead of a single transmission of the exact potential landscape, when calculating the fitness of each candidate geometry.
In contrast to the potential barrier case, the mass barriers in Fig. \ref{fig5}(b) struggle to implement a band pass filter, with particularly poor suppression of high energy transmissions. 
This is somewhat expected, as mass terms tend to suppress all transmission within their associated band gaps, so that a mass large enough to block transmissions at high energies would also quench the desired transmission in the pass range. 

Many electron optic applications involve controlled deflection of an electron beam~\cite{ghosh2009electron, chen2016electron}, so it is also important that the procedure be able to optimize structures for angle-dependent target transmissions.
Fig.~\ref{fig5}(c) shows an example of our approach to the design of a simple collimator for a single energy $E=60$ meV using mass barriers, where the target is a step at $\theta = 0.5$ rad. 
We note that since the system is translationally invariant in the $y$ direction, this amounts to optimizing transmission in the range $-0.5 \,\textrm{rad} < \theta < 0.5 \,\textrm{rad}$.
This target is achieved successfully by the $N=20$ and $N=50$ systems.
The role of the regularization term in producing smoother and simpler barrier configurations with less sharp discontinuities is particularly clear for the optimized structures in this case. 

The targets to date have been inspired by electron optic scenarios, where the incoming electrons are assumed to have a well-defined incident angle. 
However, it is equally applicable to more usual two-terminal electronic measurements, where the conductance of the device can be associated with an average over transmissions for all possible incident angles:
\begin{equation}
T(E) = \frac{2}{\pi} \int_{0}^{\pi/2} T(E,\theta)\,\cos\theta \, d\theta.
\end{equation}
Fig.~\ref{fig5}(d) considers an electronic band pass filter using the angle-averaged transmission in the target definition, where the blue curves here show the resulting transmissions for optimized potential barriers. 
We note that while this setup is more feasible experimentally, angle-averaged results in general tend to be less tunable than single-angle transmissions. 
This is because single-angle systems can exploit specific resonances that depend on the combination of energy and angle, whereas these can be averaged out in two-terminal measurements.

The specific results obtained for $5$ nm barrier segments in this work can easily be generalized to other system sizes. 
Within the Dirac approach used here, the exact same transmission is obtained if the length and energy scales are adjusted by reciprocal factors.
In realistic experimental systems, local gates can induce individual barriers with widths $\sim 20$ nm \cite{young2009quantum}, whereas patterned dielectric substrates regularly achieve periodicities between 20nm and 40nm and individual features as small as $8$ nm \cite{forsythe2018band, barcons2022engineering}. 
Recent work on programmable topographic landscapes \cite{lassaline2025gradient} use a $20$ x $20$ nm pixel size.
In these cases, we note that the induced potential in graphene changes faster than the physical feature size \cite{young2009quantum, huard2007transport, chaves2019electrostatics}, so that multiple $5$ nm segments can provide a better representation of the potential than a single larger segment.
Similarly, the smaller segment size allows us to model more complex systems, such as the smoother correlated barrier segments that occur when regularization is introduced, with higher resolution.

% However, we note that the $5$ nm segment size chosen here 

% We note that the $5$ nm barrier size used in this work is of a similar magnitude to, but somewhat smaller than the dimensions of typical experimental gate electrodes or dielectric features.
% Individual barriers with widths of order $20$ nm have been created with local gates \cite{young2009quantum}, whereas patterned dielectric substrates regularly achieve periodicities between 20nm and 40nm \cite{forsythe2018band, barcons2022engineering}.
% In the latter case, feature sizes as small as $8$ nm have been reported.
% Similarly, recent work on programmable topographic landscapes \cite{lassaline2025gradient} use a $20$ x $20$ nm pixel size.
% Our results can easily be scaled up to these ength scales by decreasing the relevant energy and potential scales by the same amount.
% However, we note that the induced potential in graphene changes faster than the gate or dielectric feature size \cite{young2009quantum, huard2007transport, chaves2019electrostatics}, so that four $5$ nm segments can be used to provide a better representation of the potential induced by a $20$ nm wide barrier than a single $20$ nm segment.
% Similarly, the smaller segment size allows us to model more complex systems, such as the smoother correlated barrier segments that occur when regularization is introduced, with higher resolution.

A natural extension to this work would be to consider tunable two-dimensional (2D) potential landscapes in graphene.
Many experimental works on electrostatically patterned graphene already focus on 2D arrays of potential dots~\cite{forsythe2018band, barcons2022engineering}, while recent experimental progress in the fabrication of 2D topographic landscapes \cite{lassaline2025gradient} shows that completely customized potential landscapes are possible. 
However, extending the approach in this work to two dimensions does not come without limitations. 
The lack of y-direction periodicity increases the computational cost required to evaluate each configuration, while the vastly increased parameter space will increase the difficulty to converge to a suitable solution. 
However, the optimization procedure has been shown to be effective in small arrays of electrostatic dots, where it can guide the design of electron optic components~\cite{ildarabadi2025tunable}.

Finally, the approach outlined here for monolayer graphene and relatively simple barriers can easily be extended to more complicated materials and setups. 
For example, bilayer graphene allows dual top and bottom gating~\cite{mccann2013electronic,zhan2012engineering,zhang2009direct}, which opens a band gap, similar to the mass terms discussed here, but in a far more realizable platform. 
Other 2D materials, such as phosphorene \cite{betancur2019electron} have been proposed as electron optic platforms and our approach could also help optimize devices in these systems. Heterostructures composed of different 2D material layers also provide an interesting playground where, instead of different potential regions, our approach could be used to determine the composition and size of different domains in regions in stacked or lateral heterostructures~\cite{novoselov20162d, wang2019recent}. 
This offers a wide variety of opportunities to optimize for target behaviors associated with important electronic~\cite{iannaccone2018quantum}, optoelectronic~\cite{pham20222d}, spintronic~\cite{gmitra2015graphene, khatibi2022emergence} or valleytronic~\cite{yu2020valleylayer, an2020valleyselective, khan2021electric} functionalities. 
While the differential evolution approach is perhaps best suited to quantities which can be quickly assessed using continuum models and simple transfer matrix techniques, as we have considered here, it can also be easily extended to more detailed tight-binding or ab-initio descriptions. 
This would allow aspects such as the importance of barrier direction, which is important in the limit of atomistically sharp barriers \cite{de2015velocity}, to be taken into account.
The approach can also be extended to more advanced quantum transport simulations using realistic multi-terminal device geometries, albeit with increased computational cost.

\section{Conclusion}
In conclusion, we demonstrated that differential evolution is a powerful method to optimize electronic transport behavior in 2D materials through the inverse construction of potential barrier configurations. 
Working with a test case with mass barriers and an arbitrary piecewise target transmission, we first highlighted different evolutionary strategies to speed up the optimization process.
We then discussed the three-way interaction that exists between the accuracy of the solution, the computational complexity of the optimization process, and the complexity of the associated barrier system.
As the last of these factors dictates the experimental feasibility of the system, we added an additional regularization term to the framework which allowed it to take experimental constraints into account.  
Finally, the simple approach was extended to angle-dependent and angle-averaged transmissions, and to more realistic transmission targets. 
The approach outlined in this work can have immediate applications to electronic and electron optic devices in graphene-based systems, but it can also be easily extended to a range of other materials and applications.

\begin{acknowledgments}
The authors acknowledge funding from Research Ireland under the Laureate (IRCLA/2017/322) and Frontiers for the Future (24/FFP-P/12941) programmes, and the gazelle computational facility in the School of Physical Sciences at DCU, which is supported by Intel Ireland.   
\end{acknowledgments}

\bibliography{refs.bib}% Produces the bibliography via BibTeX.

\end{document}